\documentclass[prl,showpacs,aps,twocolumn]{revtex4}
\usepackage{amsmath}
\usepackage{graphicx}
\usepackage{graphics}
\begin{document}

\title{Determining the electronic triplet-singlet transition probability in double quantum dots: Analogy with the double slit experiment}

\author{Fernando Dom\'inguez and Gloria Platero}
\affiliation{Instituto de Ciencia de Materiales de Madrid, CSIC,
Cantoblanco, Madrid 28049, Spain.}
\begin{abstract}

We apply an elementary measurement scheme to calculate 
the electronic triplet-singlet transition mediated by hyperfine interaction 
in a double quantum dot. We show how the local 
character of the hyperfine interaction and the nuclear back-action process (flip-flop) 
are crucial to cancel destructive interferences of the triplet-singlet 
transition probability. It is precisely this cancellation
 which differentiates the hyperfine interaction from an anisotropic magnetic field 
which mixes the triplet and the singlet eigenstates.

\end{abstract}
\maketitle

Experimental progress during the last two decades has opened the 
possibility to reduce
 semiconductor devices to the nanometer scale, such as quantum wires and quantum dots. 
Furthermore, coherent control of electronic transport and spin manipulation in quantum
 dots \cite{spinsinqudots} is also possible due to the long spin relaxation time 
\cite{fuj}. These two facts lead us to think of quantum dots as the minimal structures of
 computers based on quantum mechanical principles, i.e., quantum computers. Quantum 
properties such as entanglement or quantum parallelism will be used as algorithm tools
 \cite{Divicenko00}. However, there are still a lot of obstacles that separate us from
 the construction of such a computer. A major sources one comes from
 decoherence through the interaction with the environment. Special attention has been
 paid to the interaction between the nuclear and the electronic spins by means of the 
hyperfine interaction (HF) \cite{glaz, Merkulov02, LossDeco07}. 
The importance of this well-known decoherence process 
is clearly manifest in a very known system: a double quantum dot (DQD) in the spin
 blockade regime (SB)
 \cite{SBR, Johnson05, Pfund07, Koppens05, mcdonald06, Nazarov06, Levitov07}. 
There, the occupation depends on the spin degree of freedom and
 sequential transport is blocked due to the Pauli exclusion principle. 
In this way, whenever the transport is blocked, a current may arise only when
spin scattering processes such as HF interaction flips one of the electronic spins 
\cite{SBR, mcdonald06}, inducing the triplet-singlet transition ($T_{\pm1}$-$S$).

Many experiments have been performed in a lateral DQD in the SB regime. Some of them
 show a hysteretic behavior upon sweeping the magnetic field 
\cite{Pfund07, Koppens05}. Besides this hysteretic behavior, other experiments in the 
strong interdot coupling regime show how current changes radically and prominent current
 spikes appear tuning the in-plane magnetic field \cite{Koppens05}.
Motivated by these recent experiments, we have studied microscopically the 
$T_{\pm1}$-$S$ transition probability induced by the HF interaction in a lateral 
DQD. The $T_{\pm1}$-$S$ transition determines transport and serves as a basis to 
study the nuclear dynamical polarization, providinging the possibility to study 
quantitatively the current in any interdot coupling regime.

\begin{figure}[tb]
\begin{center}
\includegraphics[width=3in,clip]{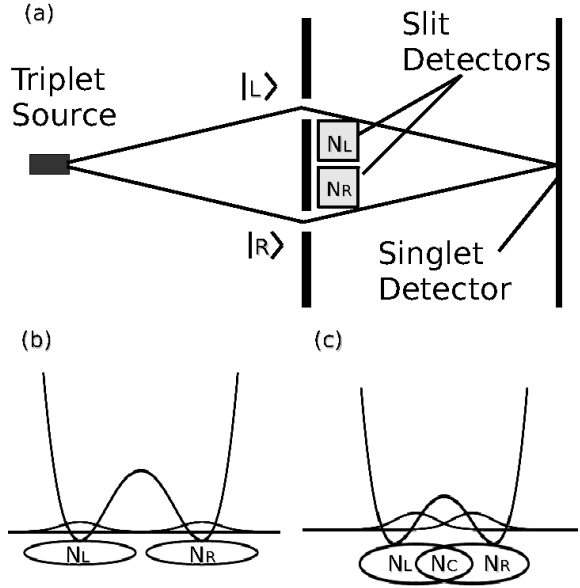}
\end{center}
\caption {\label{Fig1}\small (a) Spin Blockade regime shown in the scheme of the double slit experiment. Schematic drawn of the atomic envelope functions of a DQD in the (b) weak, (c) strong coupling regimes.}
\end{figure}

\textit{Double slit analogy}.---Before starting to calculate the transition rate, let us
 discuss some physical aspects which make the HF interaction different from 
other interactions such as the spin-orbit interaction or an anisotropic magnetic field.
 The HF interaction has two special characteristics: the
first one is its local character. 
Thus, the electronic envelope function determines the number of nuclei 
which can interact with the electronic spin. Therefore, it is natural to 
associate an ensemble of nuclear spins to each quantum dot ($N_L$ and
 $N_R$ in Figs.~1b and c). The second one is related to spin conservation. 
Whenever there is an electronic spin-flip transition ($T_{\pm1}$-$S$), 
the spin orientation of one nuclear spin localized in one of the two baths, $N_L$ or
$N_R$, is reversed. It is precisely this local
 change which allows one to detect in which of the dots the electronic spin flips. 
In analogy to the double slit experiment, we will show that the negative 
$T_{\pm1}$-$S$ interference pattern is completely destroyed when the nuclear spin 
ensembles measure exactly in which of the dots is the spin-flip produced (Fig.~1b).
 To complete our analysis, we have considered the case where some of the nuclear spins 
interact with both dots (Fig.~1c), which occurs when the electronic wave 
function is extended, i.e., strong interdot coupling. As we will see, the shared 
bath give rise to an uncertainty in the local measurement of the spin-flip, leading to
 the appearance of negative interference terms proportional to the overlap of the 
electronic wave functions.

Considering the nuclear spin bath as a slit detector is supported by the fact that
nuclear spins have no internal dynamics \cite{FeynmanIII}. Estimations of nuclear spin
dynamics, due to dipole-dipole nuclear spin interaction, suggest that time scales 
governing nuclear spin evolution $(t>100 \mathrm{ms})$ are orders
of magnitude slower than other associated with electron spin processes
\cite{Merkulov02}. 
Thus, we consider only changes of the nuclear spin states induced by the HF interaction
 with the electrons. If the internal nuclear spin dynamics were not frozen, 
one would take them into account \cite{LossDeco07}.

\textit{Transition rate}.---In our model, the eigenstates of two electrons trapped in a lateral
 DQD are obtained
 by using the Heitler-London approximation. This approximation has been widely used in
 the DQD context
\cite{LossDD98}. The electronic wave functions are described by the direct product of
 spin and orbital wave functions. The spin state is given by the known singlet-triplet
 basis, while the orbital part is composed by the displaced Fock-Darwin states
$\langle{{\bf r}}|L\rangle$ and $\langle{{\bf r}}|R\rangle$, which are the exact
 atomic states 
\begin{align}
|\phi_{\pm}\rangle=\frac{|{L(1)R(2)}\rangle\pm|{L(2)R(1)}\rangle}{\sqrt{2(1 \mp O^2)}},
\end{align}
with $O=\int d^2r \langle{L}|{\bf r}\rangle\langle{{\bf r}}|R\rangle$ corresponding
 to the overlap of the right and left orbitals. The sign $+(-)$ corresponds to
 the singlet (triplet) state, while the numbers 1 and 2 label the 
 electrons. In the spin blockade regime, electrons are found in the triplet
 states $|T_{\pm 1}\rangle$, where the two electrons have parallel spin
 polarization.

The hyperfine interaction can be seen as the scattering between electronic and
 the nuclear spin wave functions. The above mentioned characteristics  expressed by 
the contact Hamiltonian \cite{Slichter}
\begin{align}
&V_{HF}=\frac{A}{N_L}\sum_{k=1}^{N_L}\sum_{i=1,2}(S_i^+I_{L,k}^-+S_i^-I_{L,k}^++S_i^zI_{L,k}^z)\nonumber\\
&+\frac{B}{N_R}\sum_{k=1}^{N_R}\sum_{i=1,2}(S_i^+I_{R,k}^-+S_i^-I_{R,k}^++S_i^zI_{R,k}^z),
\label{eq:hiperlocali}
\end{align}
where $S_i^{\pm}$ are the raising/lowering spin operators of the electron
 ${\it i}$. The $I_{L(R),k}^{\pm}$ are the raising/lowering of the ${\it k}\mathrm{th}$
 nuclear spin operator. The subscripts $L$ and $R$ denote in which of the dots are placed
 the nuclear spins. $N_{L(R)}$ is the number of nuclear spins which interact with an 
electron when it is localized in the left (right) dot. $A$ and $B$ are the hyperfine 
interaction constants for the left and the right dot, respectively.

Terms containing the raising and the lowering operators describe the dynamic part of the
 hyperfine interaction, they are responsible for the electronic-nuclear spin-flip.
 On the other hand, the ${\it z}$-projection terms give rise to an additional Zeeman
 splitting, called Overhauser shift.

Having defined the DQD eigenbasis and the HF Hamiltonian, we are ready to study
 the decoherence produced by the spin environment in the $T_{\pm1}$-$S$ transition.
 In order to describe completely the $T_{\pm1}$-$S$ transition, we must include the
nuclear states in the initial and final wave functions \cite{FeynmanIII}. 
Thus, we obtain the transition probability rate
\begin{align}
\mathcal{P}_{T \rightarrow S}&=\sum_{k=1}^N|\langle{m_{f,k}}|\langle{S}|V_{HF}|{T_{\pm 1}}\rangle|{m_i}\rangle|^2\nonumber\\
&=\sum_{k=1}^N\langle{S}|\langle{m_{f,k}}|\rho|m_{f,k}\rangle|S\rangle,
\label{eq:probability}
\end{align}
where
$\rho=V_{HF}|{T_{\pm 1}}\rangle|{m_i}\rangle\langle{m_i}|\langle{T_{\pm 1}}|V_{HF}$.
 We have defined $N$ as the total number of nuclear spins which interact in the dots.
 The initial nuclear spin state
\begin{align}
|{m_i}\rangle=\prod_{k=1}^{N}|{\sigma_{k}}\rangle,
\label{eq:confi-nucleos}
\end{align}
is an eigenstate of $I_z$. 
Here $\sigma_{k}$ is the ${\it z}$-component of the ${\it k}\mathrm{th}$
nuclear spin. The states $|m_{f,k}\rangle$ for ${\it k}$ from 1 to
$N$, represent all possible final nuclear spin states. Depending
on the electronic initial state $|T_{\pm 1}\rangle$, it is defined
as
$|{m_{f,k}}\rangle=I_k^{\pm}|{m_{i}}\rangle$
where $I_{k}^{\pm}$ are the raising/lowering ${\it k}\mathrm{th}$ nuclear
spin operators. $|m_{f,k}\rangle$ is zero in the case where the
initial ${\it k}\mathrm{th}$ nuclear spin is parallel oriented with respect to the
electronic spins $|T_{\pm 1}\rangle$. In order to make the
discussion clearer, we restrict the calculation to the initial electronic
state $|T_{+1}\rangle$. Notice that in this case the contribution to the $T_{+1}$-S
 transition comes from the down oriented nuclear spins.

The HF interaction entangles the electronic and the nuclear wave functions.
Operating the initial state $|{\Psi}\rangle=|{T_{+1}}\rangle|{m_i}\rangle$ by
means of the HF Hamiltonian $(\ref{eq:hiperlocali})$, we obtain
\begin{widetext}
\begin{align}
V_{HF}|\Psi\rangle=\frac{1}{\sqrt{2(1+O^2)}}\biggl[|{\uparrow_1,\downarrow_2}\rangle\biggl(\frac{B}{N_R}&|{L(1)R(2)}\rangle|{M_R}\rangle-\frac{A}{N_L}|{L(2)R(1)}\rangle|{M_L}\rangle\biggr)+\nonumber\\
+&|{\downarrow_1,\uparrow_2}\rangle\left(\frac{A}{N_L}|{L(1)R(2)}\rangle|{M_L}\rangle-\frac{B}{N_R}|{L(2)R(1)}\rangle|{M_R}\rangle\right)\biggr],
\label{eq:medioro}
\end{align}
\end{widetext}
where
\begin{align}
|{M_{L(R)}}\rangle\equiv\sum_{k=1}^{N_{L(R)}}I_{L(R),k}^+|{m_i}\rangle,
\label{eq:interacnuc}
\end{align}
is a linear combination of nuclear states, where one of the
 nuclear states of the $L(R)$ dot has been flipped from $\downarrow$ to
$\uparrow$, i.e. $N_{R(L),\downarrow}$. 
Each component of the sum $(\ref{eq:interacnuc})$ belongs to the
final state basis $\{|m_{f,k}\rangle\}$. $|{M_{L(R)}}\rangle$ contains as many terms
as nuclear spins $\downarrow$ interact with an electron localized in the dot
$L(R)$.

Using eq.~$(\ref{eq:medioro})$ we can calculate directly the product
$\rho=V_{HF}|{\Psi}\rangle\langle{\Psi}|V_{HF}$. Finally we have to carry
out the projection on the final nuclear (nuclear trace out) and electronic (singlet)
 states. Before presenting the general results, it is convenient to evaluate first scalar
 products involved in the nuclear trace
\begin{align}
\sum_{k=1}^N\langle{m_{f,k}}|{M_{i}}\rangle\langle{M_{j}}|{m_{f,k}}\rangle=\langle{M_{i}}|{M_{j}}\rangle,
\label{eq:nuclgeneral}
\end{align}
for $i$ and $j$ equal to $L$ and $R$. First of
all, we evaluate the case $i=j$. Due to orthogonality, the 
projection of each component of state $(\ref{eq:interacnuc})$ contributes to the total 
scalar product with unity, if two equal components are projected, and
 zero otherwise. With this in mind it is easy to calculate
\begin{align}
\langle{M_{L(R)}}|{M_{L(R)}}\rangle=N_{L(R),\downarrow}.
\label{eq:ni}
\end{align}

On the other hand, the meaning of the scalar product with $i\not=j$ is more subtle.
 If we had imagined $|{M_L}\rangle$ and $|{M_R}\rangle$ as two independent spaces,
 we would say that no term survives due to orthogonality conditions. In principle, 
this argument looks to be reasonable, but care must be taken due to the existence of an 
overlap between the atomic wave functions $O^2$. The existence of such an overlap 
implies that both electrons can interact with a common ensemble of nuclear spins at the
 same time, see Fig.~1(c). Mathematically this is reflected by the fact that 
$|M_L\rangle$ and $|M_R\rangle$ contain a common nuclei ensemble
\begin{align}
\langle{M_{L}}|{M_{R}}\rangle=N_{C,\downarrow},
\label{eq:nc}
\end{align}
where $N_{C,\downarrow}$ is the number of nuclei that can be
flipped by both electrons placed in different dots. Obviously this number is 
proportional to the overlap. It must be noted
that $N_{i,\downarrow}$ refers to those nuclei
that can interact only with the electron localized at dot ${\it
i}$ and others which can interact with both of them. Thinking
in terms of detectors, we would say that the existence of the shared
bath gives rise to an uncertainty in the localization of the electronic 
spin-flip. Thus, the higher $N_{C,\downarrow}$, the less
reliable the spin-flip detectors and the more pronounced the negative
interference pattern.

Finally, we replace the obtained expressions $(\ref{eq:medioro})$, $(\ref{eq:ni})$,
 $(\ref{eq:nc})$ into $(\ref{eq:probability})$, yielding
\begin{align}
\mathcal{P}_{T\rightarrow S}=D\biggl(B^2\frac{N_{R,\downarrow}}{N_{R}^2}+A^2\frac{N_{L,\downarrow}}{N_{L}^2}-2AB\frac{N_{C,\downarrow}}{N_{L}N_R}\biggr),
\label{eq:general}
\end{align}
where we have used $D=\left(1+O^2\right)/\left(1-O^2\right)$. Equation 
$(\ref{eq:general})$ is the main result of our work. It is composed by three
 terms, the first two arise due to the contribution of the nuclear spins of each 
dot which are able to flip the electronic spin, while the third one arises due to the 
uncertainty in the measurement of the spin-flip position $(N_{C,\downarrow})$, caused
 by the overlap of the electronic wave functions. We observe a change 
 the tunnel interdot on the interference pattern. In the weak-coupling regime, the 
interference tends to zero since the overlap is negligible $(O^2\rightarrow0)$. In this
 case, nuclear detectors are perfectly reliable and thus the interference term of 
eq.~$(\ref{eq:general})$ is cancelled \cite{mcdonald06}. On the other hand, in 
the strong-coupling regime the overlap is not negligible and an uncertainty in the 
spin-flip position arises and leads to the appearance of a negative interference pattern.
 It must be noted the fundamental difference between the HF interaction and the effect 
of an inhomogeneous magnetic field, in the case of the inhomogeneous magnetic field 
the interference pattern holds, leading to a probability which depends on the difference
 between the in-plane effective magnetic fields of each dot \cite{Nazarov06}.

Let us extend our analysis to the transition rates between the 
triplet states $T_{\pm1}$ and $T_{0}$. It yields a similar expression as 
$(\ref{eq:general})$, except for $D$ which becomes one, and the negative sign of the
 interference pattern which becomes positive. On the other hand, the states 
$T_0$ and $S$ are mixed due to the difference between the Zeeman splittings 
within each dot, i.e., due to the magnetic field anisotropy \cite{Sousa01}.

\begin{figure}[tb]
\begin{center}
\includegraphics[width=3in,clip]{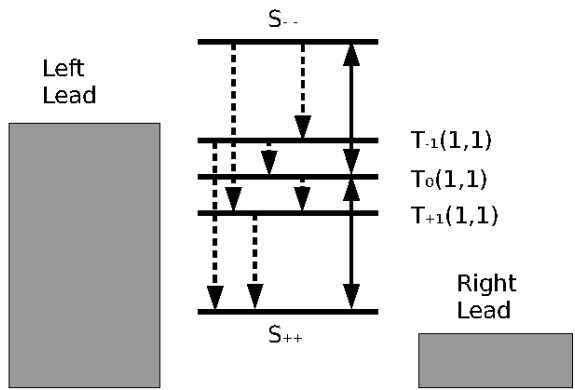}
\end{center}
\caption {\label{Fig2}\small Transport window scheme of a DQD in the sharp transport regime. The dashed arrows represent spin-flip phonon assisted transitions, while the double arrows represent coherent couplings due to the inhomogeneous Overhauser field.}
\end{figure}

\textit{Current}---In order to show how the obtained transition rate 
$(\ref{eq:general})$ determines a measurable quantity as the current, we analyze 
transport in the strong interdot coupling regime by means of a simple model. 
We focus on the following transport 
configuration: zero detuning, low in-plane magnetic field and strong interdot coupling, 
we consider the schematic picture of the different spin-flip transitions depicted in
 Fig.~2 \cite{Koppens05}. At this 
experimental conditions, the energy difference between the T(1,1) and S(1,1) 
($T_0$ and $T_{\pm1}$) states is larger than the nuclear Zeeman 
splitting. Therefore, at low temperatures only transitions $T_{\pm1}$-$S$ and 
$T_{\pm1}$-$T_0$ involving phonon emission are efficient \cite{Nazarov01}.

We calculate the stationary current through the DQD based in a model 
presented in reference \cite{Nazarov06}. Within a density matrix formalism,
 considering the electron reservoirs within Markov approximation. 
The system involves seven diagonal matrix elements: three triplet 
states $T$(1,1), two singlet states $S_{++}=\frac{1}{\sqrt{2}}\left(S(1,1)+S(0,2)\right)$
 and $S_{--}=\frac{1}{\sqrt{2}}\left(S(1,1)-S(0,2)\right)$ and two single 
occupied states \cite{nota}. Additionally, it involves six non-diagonal matrix elements, 
corresponding to the coherences between the singlet states $S_{++}$, $S_{--}$ and the 
triplet $T_0$(1,1) 
states, which are mixed by the anisotropy of the Overhauser field $(\Delta B_z)$.
 We calculate the stationary current making the time derivatives equal to zero.
 Aiming at simplicity, the triplet states T(1,1) and the extended singlet state 
$S_{++}$ are coupled to the left lead while $S_{++}$ and 
$S_{--}$ are coupled to the right, by means of the coupling constant
 $\Gamma$. The current is proportional to the occupation of the state S(0,2) and can 
be calculated analytically for the general case. The general solution is quite lengthy 
but it can be simplified assuming that $\Delta B_z\ll E_S-E_T$ and that the transition
 rate $\Gamma$ is orders of magnitude higher than the spin-flip rates, yielding
\begin{align}
I=\frac{7 \beta \delta(\alpha+\delta)}{6\beta \delta+2\delta(3\delta+2\alpha)+2\alpha\beta},
\label{eq:corriente}
\end{align}
where $\alpha$ and $\beta$ represent the inelastic transition rates 
$T_{\pm1}$-$S$, while $\delta$ represents the rates $T_{\pm1}$-$T_0$. 
Obviously this simple model does not attempt to explain quantitative experimental 
evidences \cite{Koppens05}, but it is illustrative in order to show how the current
 is governed mainly by the transition rates $T_{\pm1}$-$S$ and 
$T_{\pm1}$-$T_0$ $(\ref{eq:general})$. To obtain a more detailed model 
one has to study the time evolution accounting for the dynamical polarization of the 
nuclear spin ensembles, which is responsible for current bistability among other 
non-linear effects \cite{mcdonald06}.

\textit{Conclusions}.---We have presented a microscopic model to describe the triplet-singlet
 and triplet-triplet transition probabilities mediated by the HF interaction in 
a DQD. We have stressed the importance of the local character and the 
nuclear flip-flop process of the HF interaction. These characteristics lead to a
partial cancellation of the interference pattern, which can be intuitively seen by means
 of an analogy between the triplet-singlet transition and the double-slit experiment. 
With this picture in mind, we have shown the fundamental difference between the 
transition mediated by the hyperfine interaction and an anisotropic magnetic field.
 The transition under study turned out to be relevant 
in the spin blockade regime. The obtained results will serve as a basis to 
study transport accounting for the nuclear spin dynamical polarization and will open the 
possibility to explain experiments covering different tunneling coupling regimes.

We like to thank C. L\'opez-Mon\'is and J. I\~narrea for fruitful
 discussions and S. Kohler for helpful comments on the paper. 
This work has been supported by the Spanish project MAT2008-02626. F. D. 
acknowledges MEC (Spain) for financial support through the FPI predoctoral
 grant.


\begin{thebibliography}{16}

\bibitem{spinsinqudots}
R. Hanson {\it et al.}, Rev. Mod. Phys. {\bf 79}, 1217 (2007).
\bibitem{fuj}
 T. Fujisawa, D. G. Austing, Y. Tokura, Nature 419, 278 (2002); A. C. Johnson {\it et al.}, Nature 435, 925 (2005).
\bibitem{Divicenko00}
A. Ekert and R. Josza, Rev. Mod. Phys. {\bf 68}, 733, (1996); C. H. Bennett and D. P. DiVicenzo, Nature (London) {\bf 404}, 247 (2000).
\bibitem{glaz}
A. Khaetskii, D. Loss and L. Glazman, Phys. Rev. B, {\bf 67}, 195329
(2003).
\bibitem{Merkulov02}
I. A. Merkulov, A. Efros and M. Rosen, Phys. Rev. B {\bf 65}, 205309, (2002).
\bibitem{LossDeco07}
W. A. Coish {\it et al.}, J. Appl. Phys. {\bf 101}, 081715 (2007); A. V. Khaetskii, D. Loss and L. Glazman. Phys. Rev. Lett. {\bf 88}, 186802 (2002); A. Relao, J. Dukelsky and R. A. Molina, Phys. Rev. E {\bf 76},
046223, (2007).
\bibitem{SBR}
K. Ono {\it et al.}, Science {\bf 297} 1313 (2002); A. C. Johnson {\it et al.}, Phys. Rev. B {\bf 72} 165308 (2005).
\bibitem{mcdonald06}
J. I\~narrea, G. Platero and A. H. MacDonald, Phys. Rev. B {\bf 76},
085329, (2007); J. I\~narrea, C. L\'opez-Mon\'is, A. H. MacDonald and G. Platero, Appl. Phys. Lett., {\bf 91}, 252112 (2007); J. I\~narrea, C. L\'opez-Mon\'is and G. Platero, Appl. Phys. Lett., {\bf 94}, 252106 (2009).
\bibitem{Johnson05}
A. C. Johnson {\it et al.}, Nature {\bf 435}, 925 (2005).
\bibitem{Pfund07}
A. Pfund {\it et al.}, Phys. Rev. Lett. {\bf 99}, 036801, (2007).
\bibitem{Koppens05}
F. H. L. Koppens {\it et al.}, Science {\bf 309}, 1346 (2005).
\bibitem{Nazarov06}
O. N. Jouravlev and Y. Nazarov, Phys. Rev. Lett. {\bf 96}, 176804 (2006).
\bibitem{Levitov07}
M. S. Rudner and L. S. Levitov, Phys. Rev. Lett. {\bf 99}, 036602, (2007).
\bibitem{FeynmanIII}
R. P. Feynman, R. B. Leighton and M. Sands, {\it The Feynman Lecture on Physics 3}, p. 3-7, Addison-Wesley (1963).
\bibitem{LossDD98}
G. Burkard, D. Loss and D. P. DiVicenzo, Phys. Rev. B {\bf 59}, 2070 (1998); R. de Sousa, X. Hu and S. Das Sarma, Phys. Rev. A {\bf 64}, 042307 (2001).
\bibitem{Slichter}
C. P. Slichter, {\it Principles of Magnetic Resonance}, (Springer-Verlag, Berlin, 1989).
\bibitem{nota}
The numbers $(n,n')$ specify the extra number of electrons in the left and right dot respectively.
\bibitem{Sousa01}
R. de Sousa, X. Hu and S. Das Sarma, Phys. Rev. Lett., {\bf 86}, 918 (2001).
\bibitem{Nazarov01}
S. I. Erlingsson, Y. V. Nazarov and V. Fal'ko, Phys. Rev, B, {\bf 64}, 195306 (2001).
\end{thebibliography}
\end{document}